*Snowmass 2013 White Paper*
*Novel Compact Accelerator-Based Neutron and Gamma Sources for Future Detector Calibration*

G. Jennings, C. Sanzeni, D.R. Winn*
Fairfield University, Fairfield CT 06824
**Corresponding Author* winn@fairfield.edu 1+203.984.3993

*Abstract:* Novel ultra-compact, electrically switchable, time-structured/pulsed, ~1-14 MeV-level neutron and photon generators have application embedded into large detector systems, especially calorimeters, for energy and operational calibration. The small sizes are applicable to permanent in-situ deployment, or able to be conveniently inserted into large high energy physics detector systems. For bench- testing of prototypes, or for detector module production testing, these compact n and gamma generators offer advantages.

*Introduction:*
Calibration of large experiments is an ongoing issue in HEP experimentation. Convenient sources of radiation which could be turned on or off remotely, with temporal modulation, and when off have little or no radiation, offer distinct advantages, especially over radioactive sources which must be mechanically inserted (for example, the source wires used in the CMS hadron calorimeter), and which have no timing properties. Much larger neutron generators have been used in off-line calorimeter calibration[1],[2]. We review two novel devices which have application in the energy, intensity and cosmic frontiers for high energy physics experiments.

*Ultra-Compact Neutron Accelerator-Based Source:* Sandia National Labs(SNL) has announced a miniaturized neutron generator dubbed the "Neutristor", using ceramics with an ion drift region of a few $mm^3$[3]. The DT/DD neutristor operates in CW (100's hours) pulsed operation, and produces up to ~$10^{4-5}$ 14.1 MeV DT n/s, with a lower(x0.01) 2.5 MeV DD neutron rate. The deuterium-tritium base reaction pulsed neutron generator is packaged in a flat computer chip shape of 1.54 cm(0.600 in) wide by 3.175 cm(1.25 in) long by 3mm thick has demonstrated $10^3$ neutrons per pulse (14 MeV) in a 500ns pulse.

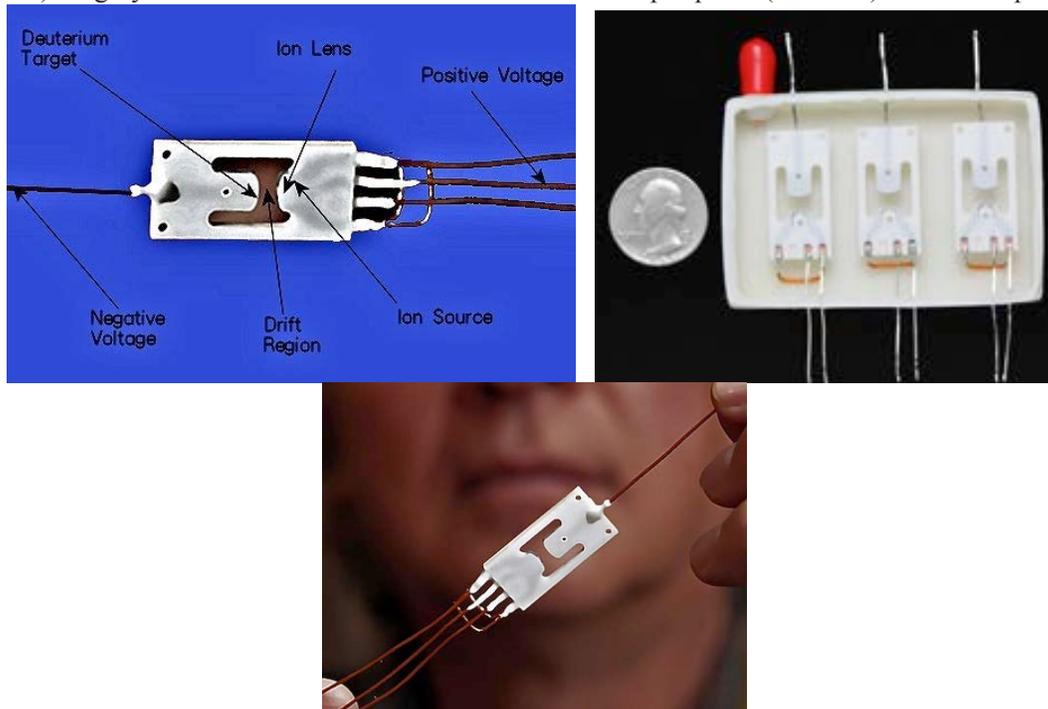

*Figs 1:* The Sandia Neutristor neutron-generators – size 1.25"x0.6"x0.13" - 1000n per 500ns pulses.

The Neutristor neutron generator is based on a deuterium ion beam accelerated to impact a tritium-loaded target. The accelerating voltage is in the 15- to 20-kV range with a 3-mm gap, and the ion beam is shaped by using a lens design to produce a flat ion beam that conforms to the flat rectangular target. The ion source is a simple surface-mounted deuterium-filled titanium film with a fused gap that operates at a current-voltage design to release the deuterium during a pulselength of ≤1 μs. The 4-terminal neutristor could easily be slotted into or assembled into large calorimeters or other detector systems, depending on the application; for example, switched on between fills in the LHC. The estimated production cost for neutristors is in the neighborhood of $2,000US and are assumed to be disposable if necessary. Sandia is working on neutristors ~2 orders of magnitude smaller, fabricated using MEMS.

*Compact 11.7 MeV Gamma Source:* Drs. Ka-Ngo Leung of Lawrence Berkeley Lab(LBNL) and Arlyn Antolak of Sandia Lab(SNL) have invented a novel neutron/gamma generator in which the electrostatic acceleration column is eliminated via a plasma generator[4].

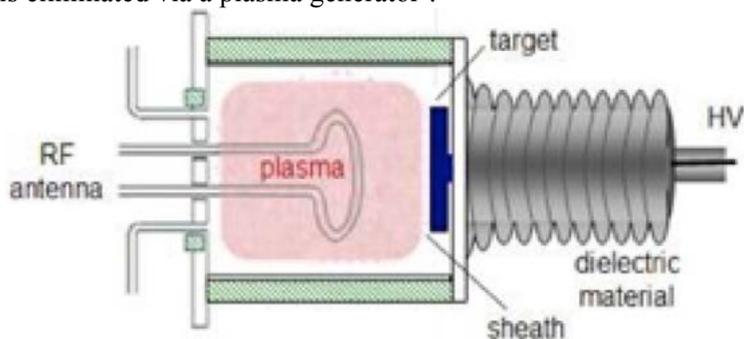

*Fig 2:* Schematic of a compact plama driven n/gamma accelerator, *5 cm long x 1.5 cm diameter* for generating $10^6$ 11.6 MeV gamma/s.

Due to the elimination of the electrostatic accelerator column, the Berkeley neutron/gamma generator can be operated at higher pressures than previous D-T or similar neutron tubes, so it is ideal for sealed tube operation. As a result, the generator can be very compact in size—as small as 1.5 centimeters in diameter and 5 centimeters in length. When operated with a $LaB_6$ target the $p+B^{11}$ gamma source generates up to ~1 million 11.7 MeV gamma per s, via the $^{11}B(p, \gamma)^{12}C$ reaction. D-D, D-T, or T-T neutrons can be generated at high current density. The generator consists of a sealed chamber, an RF antenna that can be placed either inside the ion source chamber or on the external surface of the chamber wall, a gas source, the target, a power source, and permanent magnets surrounding the target to suppress electron-generated x-rays. After a plasma is formed, a series of high voltage pulses are applied to the target, forming a plasma sheath that serves as an accelerating gap. For creating either a neutron beam or a gamma beam, the design is essentially the same, with changes in appropriate source gases and target material. Other gamma generators can be converted from neutron generators via the reaction $^{19}F(n,\alpha)^{16}N$, followed $^{16}N$ decay yielding a 6.13 MeV photon.

**SUMMARY:** Recent developments in ultra-compact, time-resolvable, sealed tube neutron and gamma generators are applicable to in-situ detector calibration, particularly calorimeters, and to testing prototypes and production modules, and merit development, especially for shorter than 500ns pulse-widths.